\pdfoutput=1
\documentclass[journal]{IEEEtran}
\usepackage[explicit]{titlesec}
\usepackage{graphicx}
\usepackage{color, soul}
\usepackage{epstopdf}
\usepackage{subfigure} 
\usepackage{algorithmic}
\usepackage{amsmath}
\usepackage{amsthm}
\usepackage{cite}

\usepackage{bbm}
\usepackage[backref]{hyperref}
\usepackage{amssymb}

\newcommand{\RNum}[1]{\uppercase\expandafter{\romannumeral #1\relax}}

\usepackage{stfloats}

\usepackage{color}
\usepackage{tikz,xcolor,hyperref}

\definecolor{lime}{HTML}{A6CE39}

\titlespacing{\section}{0pt}{1.2ex plus .0ex minus .0ex}{.3ex plus .0ex}
\titlespacing{\subsection}{0pt}{1.2ex plus .0ex minus .0ex}{.3ex plus .0ex}

\makeatletter
\DeclareRobustCommand{\orcidicon}{%
	\begin{tikzpicture}
	\draw[lime, fill=lime] (0,0) 
	circle [radius=0.16] 
	node[white] {{\fontfamily{qag}\selectfont \tiny ID}};    \draw[white, fill=white] (-0.0625,0.095) 
	circle [radius=0.007];    \end{tikzpicture}
	\hspace{-2mm}}
\foreach \x in {A, ..., Z}{%
\expandafter\xdef\csname orcid\x\endcsname{\noexpand\href{https://orcid.org/\csname orcidauthor\x\endcsname}{\noexpand\orcidicon}}
}

\newcommand*\bigcdot{\mathpalette\bigcdot@{.5}}
\newcommand*\bigcdot@[2]{\mathbin{\vcenter{\hbox{\scalebox{#2}{$\m@th#1\bullet$}}}}}
\makeatother

\usepackage[linesnumbered,ruled,vlined]{algorithm2e}

\begin{document}
\title{Deep Reinforcement Learning-Assisted Age-optimal Transmission Policy for HARQ-aided NOMA Networks}
\author{Kunpeng Liu\orcidA{},
	Aimin Li\orcidB{}, 
	\emph{Graduate Student Member, IEEE,}
	Shaohua Wu\orcidC{}, 
	\emph{Member, IEEE}
	\thanks{
		This work has been supported in part by the Guangdong Basic and Applied Basic Research Foundation under Grant no. 2022B1515120002, and in part by the National Natural Science Foundation of China under Grant no. 61871147.
		
		K. Liu and A. Li are with the Department of Electronics and Information Engineering, Harbin Institute of Technology (Shenzhen), Shenzhen 518055, China (e-mail: 190210232@stu.hit.edu.cn; liaimin@stu.hit.edu.cn).
		
		S. Wu is with the Guangdong Provincial Key Laboratory of Aerospace Communication and Networking Technology, Harbin Institute of Technology (Shenzhen), Shenzhen 518055, China, and also with the Peng Cheng Laboratory, Shenzhen 518055, China (e-mail: hitwush@hit.edu.cn).

	}
}

\maketitle
\begin{abstract}
The recent interweaving of AI-6G technologies has sparked extensive research interest in further enhancing reliable and timely communications. \emph{Age of Information} (AoI), as a novel and integrated metric implying the intricate trade-offs among reliability, latency, and update frequency, has been well-researched since its conception. This paper contributes new results in this area by employing a Deep Reinforcement Learning (DRL) approach to intelligently decide how to allocate power resources and when to retransmit in a \emph{freshness-sensitive} downlink multi-user Hybrid Automatic Repeat reQuest with Chase Combining (HARQ-CC) aided Non-Orthogonal Multiple Access (NOMA) network. Specifically, an AoI minimization problem is formulated as a Markov Decision Process (MDP) problem. Then, to achieve deterministic, age-optimal, and intelligent power allocations and retransmission decisions, the Double-Dueling-Deep Q Network (DQN) is adopted. Furthermore, a more flexible retransmission scheme, referred to as Retransmit-At-Will scheme, is proposed to further facilitate the timeliness of the HARQ-aided NOMA network. Simulation results verify the superiority of the proposed intelligent scheme and demonstrate the threshold structure of the retransmission policy. Also, answers to whether user pairing is necessary are discussed by extensive simulation results.
\end{abstract} 


\IEEEpeerreviewmaketitle

\section{Introduction} 
With the rapid development of the 5th Generation  Mobile Communication (5G), more and more Internet of Things (IoT) devices are accessed to the networks \cite{xia2020maritime}. In the IoT application scenarios, such as automatic driving, smart factory, and smart healthcare, etc., the timeliness of information plays a significant role in enabling timely, accurate, and effective decision-making. In this regard, it is imperative to deliver \emph{fresh} status updates in a timely and reliable manner, since the \emph{stale} one usually contains little value. 

To measure the timeliness of information, \emph{Age of Information} (AoI) is first proposed in \cite{6195689} for automated driving applications. AoI is defined as the time elapsed since the generation of the latest status which is received successfully. Different from the traditional metric in the 5G ultra-reliable-low-latency (URLLC), such as latency and throughput, AoI contemplates an intricate trade-off among reliability, latency, and update frequency. As such, AoI is also known as a type of semantic metrics, and has attracted massive interest from government agencies, industry, and academia.


With the massive access of IoT devices, the research of high timeliness access technique becomes a hot topic. Non-orthogonal multiple access (NOMA) is considered as a promising access technique for the future network due to its high spectral efficiency \cite{7842433}. In \cite{9417122} and \cite{wang2020minimizing}, the authors have shed light on the potential superior AoI performance in the high signal-to-noise ratio (SNR) regime. However, in the low SNR regime, the age performance of NOMA system inexorably faces a bottleneck. 

To this end, HARQ is introduced in \cite{1091865} to ensure reliability in the low SNR regime by achieving retransmission-driven time diversity. Up to this point, the HARQ-aided NOMA network has demonstrated its superiority in both throughput and latency \cite{ghanami2020performance} and \cite{marasinghe2020block}, while the age performance of HARQ-aided NOMA network is seldom investigated. An exception is our previous work in \cite{wu2022minimizing}, which analyzes the average blocklength error ratio (BLER) of users in the HARQ-aided NOMA network, and provides an optimal transmission policy by formulating an MDP problem. However, the theory and analysis in \cite{wu2022minimizing} limited the number of users and the maximum transmission round to 2 only, which hinders flexible and timely system design. 

In the multi-user NOMA network, user paring is a commonly-implemented technique to simplify the transmission mode. \cite{8408563} has demonstrated that in terms of throughput performance, the optimal pairing policy manifests a strongest-weakest symmetrical structure, i.e., the user with the best channel condition is paired with the user with the worst channel condition sequentially, so as to accomplish the pairing of all users. Nevertheless, the insight on whether user pairing will benefit the timeliness performance of the NOMA-based network remains an open issue. 

Besides, DRL algorithms are widely used in wireless communication to solve high-complexity problems. Reinforcement on Federated (RoF) scheme, based on deep multi-agent reinforcement learning is proposed in \cite{9562748}. In \cite{9384272}, authors adopt TD3 algorithm to allocate the available resources for IoT.

Motivated by the above, we aim to investigate the optimal policy to minimize the average AoI in an HARQ-aided NOMA network with multiple users and unrestricted transmission rounds. Specifically, this paper achieves three-fold contributions:
\begin{itemize}
    \item We optimize the average AoI of an HARQ-aided NOMA network with multiple users and unrestricted transmission rounds. A Retransmit-At-Will policy that allows BS to retransmit flexibly is proposed, which manifests superiority in terms of AoI compared to existing works.
    \item We reformulate the age-optimal problem as an MDP. A Double-Dueling-DQN is trained to intelligently allocate power resources and retransmit at will. The solutions reveal the threshold structure of the retransmission policy.
    \item We investigate the user pairing issue in terms of AoI under the HARQ-aided NOMA network. The simulation results illustrate that user pairing can only enhance the AoI at low SNR.
    
\end{itemize}

\section{SYSTEM MODEL AND PROBLEM FORMULATION}\label{section II}
\subsection{Transmission Modle of HAQR-aided NOMA Network}
\begin{figure}
    \centering
    \includegraphics[angle=0,width=0.5\textwidth]{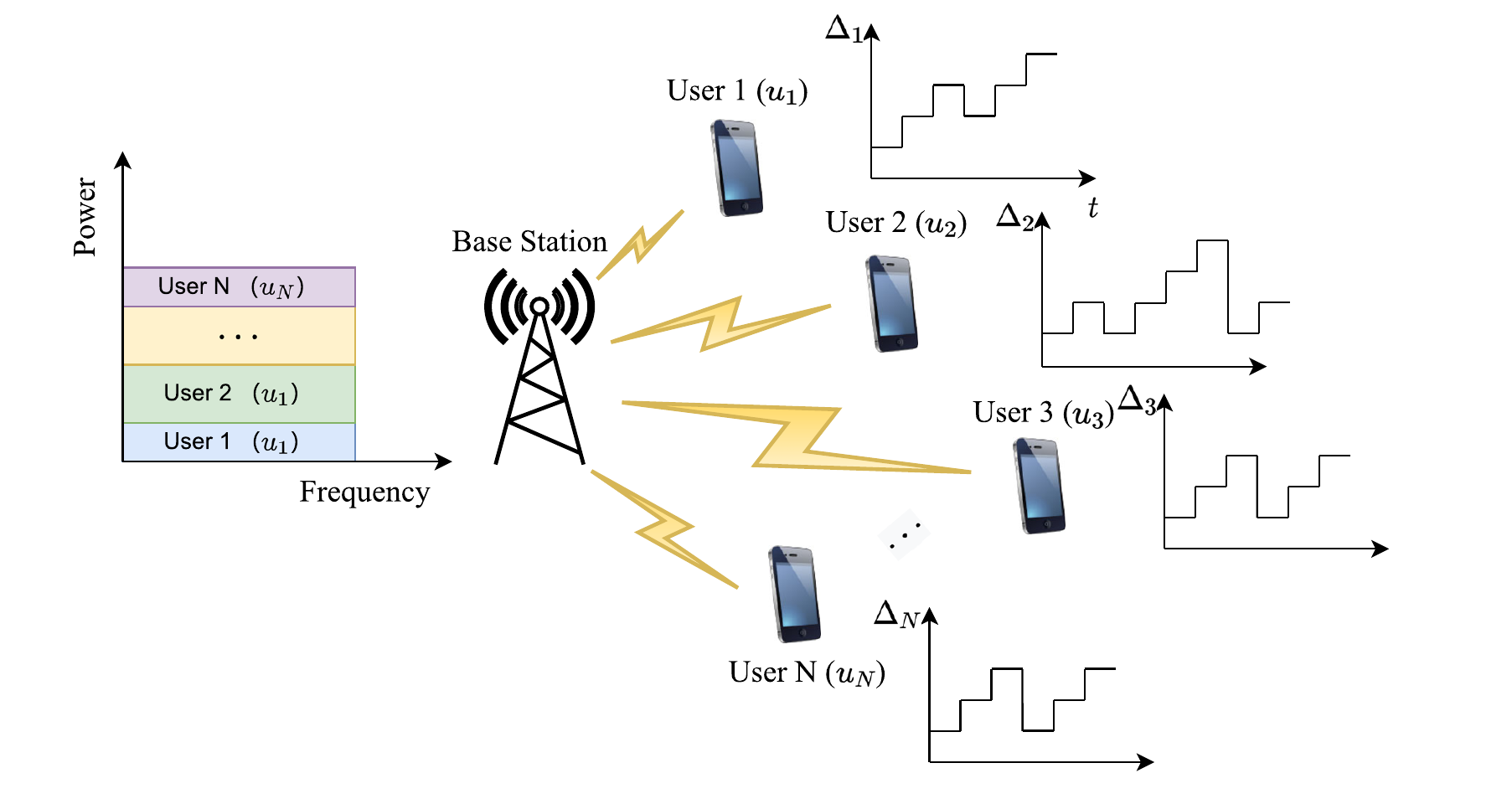}
    \caption{HARQ-aided NOMA Network Model.}
    \label{fig:1}
\end{figure}
We consider a NOMA downlink network with a BS and $N$ users $ \left(u_i,i\in\{1,\cdots,N\}\right) $ as Fig. \ref{fig:1}, the distances from users to the BS are ordered as $d_1<d_2<d_3<\cdots<d_N$. The BS sends freshness critical control commands or status updates to the users on the same shared channel. In a NOMA downlink network, the BS sends out multiple user signals by superposition coding (SC) technique in the power domain, and HARQ-CC is applied as a key technique to enhance transmission reliability by multiple retransmissions. Then, the signal sent by the BS in the $k-$th transmission can be expressed as
\begin{equation}
s_k=\sum_{i=1}^N{\sqrt{\alpha _{i,k}P}x_{i,k}}\,,
\label{eq:1}
\end{equation} 
where $P$ denotes the total power of a single transmission, $\alpha_{i,k}$ and $x_{i,k}$ denotes the power allocation coefficient and the unit energy signal of user $i$ in the $k-$th transmission. Since the receiver end of the NOMA network needs to decode superposed signals through successive interference cancellation (SIC), that is, a strong user with high channel quality needs to decode and eliminate weak users' signals iteratively, and then decodes its own signal. Generally, in order to improve the reliability of SIC, the power allocation coefficients satisfy $\alpha _{1,k}<\alpha _{2,k}<\cdots <\alpha _{N,k}$, $\sum_{i=1}^{N}{\alpha_{i,k}}=1$, which indicates that weaker users are allocated with higher power.

We consider a wireless fading channel where the signal will experience quasi-static Rayleigh fading. The perfect channel state information (CSI) is assumed to be shared with the BS. Then the channel coefficient in the $k-$th transmission from user $u_i$ to BS can be expressed as
\begin{equation}
	\label{eq:2}
h_{i,k}=\sqrt{{d_i}^{-\tau}}g\,,
\end{equation}
where $g\sim\mathcal{C} \mathcal{N} \left( 0,1 \right) $ denotes the unit Rayleigh channel coefficient and $\tau$ denotes the path loss exponent. In such a case, the signal received by $u_i$ in the $k-$th transmission is given by:
\begin{equation}
	\label{eq:3}
y_{i,k}=h_{i,k}\sum_{i=1}^N{\sqrt{\alpha _{i,k}P}x_{i,k}}+n_{i,k}\,,
\end{equation}
where $n_{i,k}$ denotes the unit complex additive white Gaussian noise (AWGN) with zero mean and variance $\sigma^2=1$.

\subsection{AoI Evolution of HARQ-aided NOMA Network}

Noticing that as the number of NOMA users increases, SIC decoding becomes more complex, which makes it harder for users to decode successfully.
To this end, we combine NOMA with HARQ-CC to ensure reliable reception of updates through retransmitting the former package. In HARQ-CC protocol, if $u_i$ decodes its message successfully, a positive acknowledgment (ACK) will be sent to the BS, otherwise, a negative acknowledgment (NACK) will be sent. Different from \cite{wu2022minimizing}, where $i$) the BS can choose whether retransmit or not only when it receives an NACK; $ii$) the BS can only transmit a new package if it receives an ACK. We propose a Retransmit-At-Will policy, which further relaxes these constraints, that is the BS can choose whether retransmit or not no matter what it receives.

We consider a time-slotted system where the continuous time is discretized into time slots\footnote{This is an extensively-implemented operation, which is also conducted in \cite{champati2019performance,krikidis2019average}}.  The BS serves all the users through the \textit{generate-at-will} model, which means the BS can generate packages for all the users at the beginning of every time slot. We adopt AoI as a metric to measure the timeliness of the information received by users. Let $\varDelta _i\left( t \right)$ denote the instantaneous AoI of $u_i$ at the time slot $t$, then if $u_i$ receives a status update successfully, $\varDelta _i\left( t \right)$ will decrease to the instantaneous AoI of this package, while $\varDelta _i\left( t \right)$ will increase by $1$ if $u_i$ fails to decode its message. We denote $t_i,t_i^{'}\in\{1,2,\cdots,T_{max}\}$ as the transmission round of $u_i$'s package at the time slot $t,t+1$ respectively, which can also represent the instantaneous AoI of $u_i$'s package, and $T_{max}$ represents the maximum transmission rounds. Then the AoI evolution of $u_i$ between two adjacent time slots can be expressed as
\begin{equation}
	\label{eq:4}
\varDelta _i\left( t+1 \right) =\begin{cases}
	\varDelta _i\left( t \right) +1,& w.p.\,\, \varepsilon _i\\
	t_{i}^{'}, & w.p.\,\, 1-\varepsilon _i\\
\end{cases}\,,
\end{equation}
where $\varepsilon _i$ denotes the BLER of $u_i$.

Noticing that if the BS retransmits the $u_i$'s package, the transmission round $t_i$ will increase by 1, and if the BS transmits a new package for $u_i$, $t_i$ will be set as 1. Therefore, we can obtain: 
		\begin{equation}
  \label{eq:05}
		t_{i}^{'}=\begin{cases}
			t_i+1,&		 \chi_i=1\\
			1, &		 \chi_i=0\\
		\end{cases},
		\end{equation}
where $\chi_i=1$ represents retransmitting the former $u_i$'s package, and $\chi_i=0$ represents transmitting a new package for $u_i$. 

\begin{figure}
    \centering
    \includegraphics[angle=0,width=0.45\textwidth]{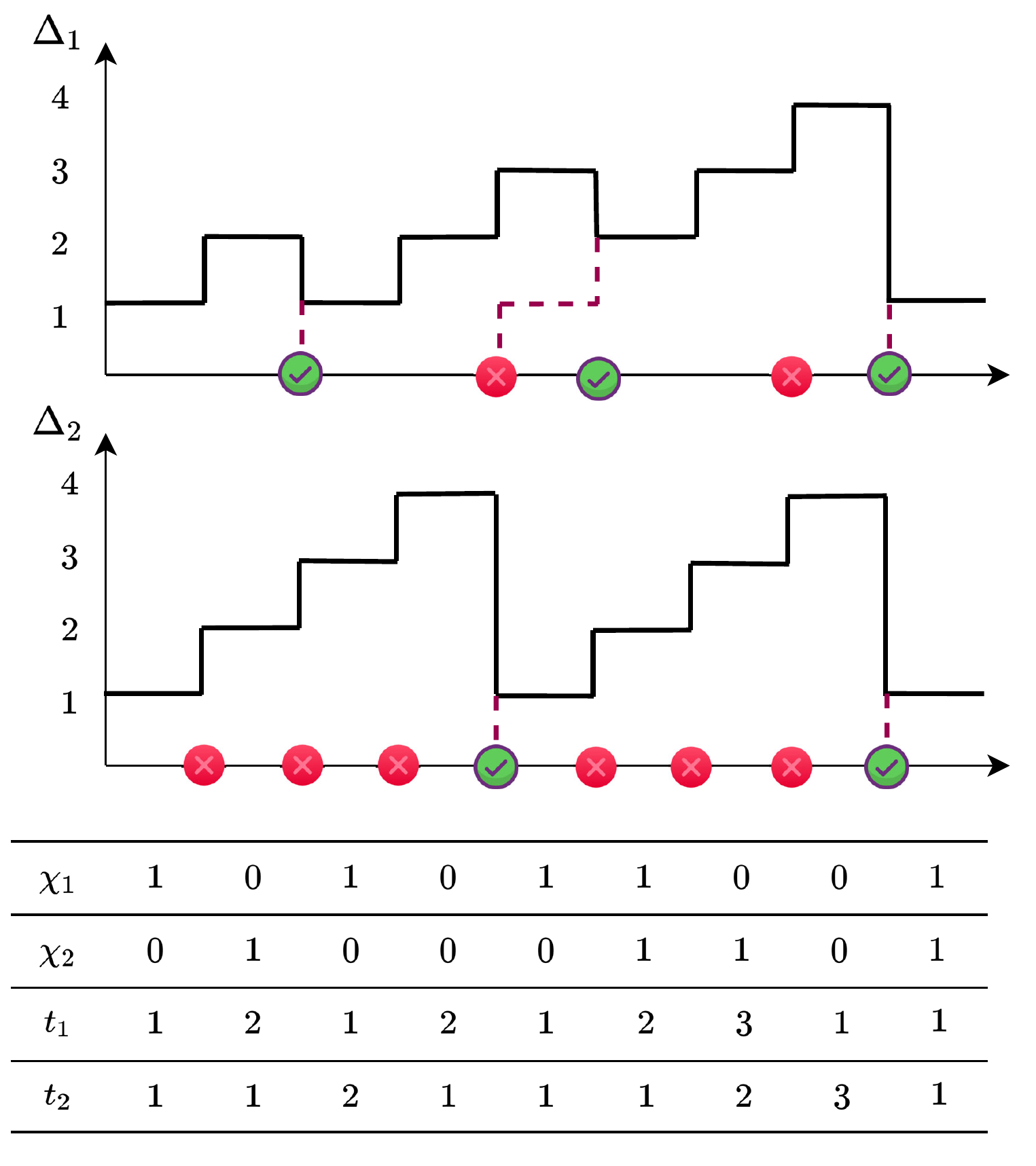}
    \caption{A two-user example of instantaneous AoI evolution.}
    \label{fig:2}
\end{figure}

Fig. \ref{fig:2} gives the AoI evolution of the HARQ-aided NOMA network with two users as an example.

\subsection{BLER analysis}
In the $k-$th transmission, only the weakest user $u_N$ can decode its message $x_{N,k}$ from $y_{N,k}$ directly. Other users $u_i (i\ne N)$ must use SIC technique to decode their message, which means $u_i$ should decode and eliminate $x_{j,k}(i\leqslant j\leqslant N)$ from $y_{i,k}$ first, and then decode its own message $x_{i,k}$. In this case, the signal-to-interference plus noise ratio (SINR) at $u_i$ for decoding $x_j$ in the $k-$th transmission can be given as:
\begin{equation}
	\label{eq:5}
\gamma_{ij}^{(k)} =\frac{\alpha _{j,k}\left| h_{i,k} \right|^2}{\sum_{n=1}^{j-1}{\alpha _{n,k}\left| h_{i,k} \right|^2+1/\rho}}\,\,(i\leqslant j\leqslant N)\,,
\end{equation}
where $\rho=P/\sigma^2$ denotes the SNR.

In the HARQ-aided NOMA network, users can store the transmitted packages. When these packages are retransmitted, the stored information can be used to achieve maximal ratio combining (MRC) decoding to improve transmission reliability. By this means, the SINR of $u_i$ for decoding the message of $u_j$ in the $t_j$-th transmission round can be given as:
\begin{equation}
	\label{eq:6}
	\gamma _{ij}\left( t_j \right) =\sum_{k=1}^{t_j}{\gamma _{ij}^{(k)}}\,,
\end{equation}
where $t_j\in{\{1,2,\cdots,T_{max}\}}$.

In \cite{yu2017performance}, the BLER of $u_i$ for decoding $u_j$'s message which has been transmitted for $t_j$ times can be approximated as:
\begin{equation}
	\label{eq:7}
\varepsilon _{ij}\left( t_j \right)\approx Q\left( \frac{C\left( \gamma _{ij}\left( t_j \right) \right) -\frac{N}{m}}{\sqrt{V\left( \gamma _{ij}\left( t_j \right) \right) /m}} \right)\,,
\end{equation}
where $N$ denotes the length of message sent to $u_j$, $m$ denotes the number of encoded symbols, $C\left( \gamma _{ij}\left( t_j \right) \right)=\log _2\left( 1+\rho \right) $ is the channel capacity, and $V\left( \gamma _{ij}\left( t_j \right) \right)$ is the channel dispersion, given as $V\left( \gamma _{ij}\left( t_j \right) \right)=(1-1/(1+\gamma _{ij}\left( t_j \right) )^2)(\log_2e)^2$.

Denote a vector $\boldsymbol{\overset{\sim}{\alpha_k}}\triangleq \left( \alpha _{1,k},\alpha _{2,k},\cdots ,\alpha _{N,k} \right) $ as the allocated power coefficients for $N$ users at the $k-$th transmission. According to equation (\ref{eq:5})(\ref{eq:6})(\ref{eq:7}), when given a fixed $N$ and $m$, $\varepsilon _{ij}\left( t_j \right)$ is equivalent to a function of $t_j$ vectors. In this case, $\varepsilon_{ij}(t_j)$ can be expressed as:
\begin{equation}
	\label{eq:8}
\varepsilon _{ij}\left( t_j \right)\approx\psi _{ij}\left( \boldsymbol{\overset{\sim}{\alpha _1}},\boldsymbol{\overset{\sim}{\alpha _2}},\cdots ,\boldsymbol{\overset{\sim}{\alpha _{t_j}}} \right)\,, 
\end{equation}

 According to the principle of SIC, only when the user successfully decodes and eliminates other interference signals, and then successfully decodes its own signal, it is a successful decoding process. In this case, the BLER of $u_i$ can be given as:
\begin{equation}
	\label{eq:9}
\varepsilon _i=1-\prod_{j=i}^N{\left( 1-\varepsilon _{ij}\left( t_j \right) \right)}\,,
\end{equation}
which means the BLER of $u_i$ is associated with the transmission round of $u_j$'s signal $t_j (i\leqslant j\leqslant N)$ and the corresponding power allocation coefficient vector $ \boldsymbol{\overset{\sim}{\alpha _k}}$.

\subsection{Problem Formulation}
In the HARQ-aided NOMA network, retransmissions can enhance the reliability but cause information aging. Besides, how to allocate power resources has a significant impact on the system timeliness. Therefore, it's necessary for the BS to adjust transmission policy $\pi$ according to the state of the network to enhance the timeliness. 

We use the expected weighted sum AoI of all the users to measure the AoI performance of the HARQ-aided NOMA network. To minimize the expected weighted sum AoI, the problem can be formulated as follows:
\begin{equation}
	\label{eq:10}
\mathop{\arg\min}\limits_{\pi}\bar{\varDelta}=\lim_{T_{slot}\rightarrow \infty} \frac{1}{T_{slot}}\sum_{t=1}^{T_{slot}}{\sum_{i=1}^N{\mathbb{E} \left[ \omega _i\varDelta _{i}^{\pi}\left( t \right) \right]}}\,,
\end{equation}
where $\omega_i$ denotes the weight of $u_i$. The transmission policy $\pi$ contains two aspects: $i$) the power allocation scheme $\boldsymbol{\overset{\sim}{\alpha}}$ at every time slot; $ii$) when and whether to retransmit the former package.

\section{MDP FORMULATION}\label{sectionIII}
To obtain the age-optimal transmission policy for the HARQ-aided NOMA network, we reformulate the problem (\ref{eq:10}) as an MDP problem, which is characterized by a quaternary tuple $\left\{ \mathcal{S} ,\mathcal{A} ,\mathcal{P} ,r \right\} $, where $\mathcal{S} ,\mathcal{A} ,\mathcal{P} ,r$  denotes the state space, action space, state-action transition probability matrix, and the reward respectively. The elements of the MDP for HARQ-aided NOMA network are described specifically as follows:

\begin{itemize}
	\item \textbf{State Space:} We denote $s_t $ as the state of the network at the time slot $t$, which is characterized by a tuple $ s_t \triangleq \left( \boldsymbol{\overset{\sim}{\alpha _1}},\boldsymbol{\overset{\sim}{\alpha _2}},\cdots ,\boldsymbol{\overset{\sim}{\alpha _{T_{max}-1}}},\boldsymbol{\mathcal{T}},\boldsymbol{\varDelta} \right) $, where
	\begin{enumerate}
		\item The coefficient vector $\boldsymbol{\overset{\sim}{\alpha _i}}\triangleq(\alpha_{1,i},\alpha_{2,i},\cdots,\alpha_{N,i})$ denotes the $i-$th power allocation coefficient for all the $N$ users which is stored in the buffer. We discretize the power into $M$ levels, and the power coefficient can only take value from a discrete set, that is $\alpha_{i,k}\in\{\frac{1}{M},\frac{2}{M},...,\frac{M-1}{M}\}$. Due to the computational complexity, the case $\alpha_{i,k}=1$ which means BS serves only one user is not taken into consideration.
		\item The vector $\boldsymbol{\mathcal{T}}\triangleq(t_1,t_2,\cdots,t_N)$, where $t_j$ denotes the transmission round of $u_j$'s signal at the time slot $t$, with $t_j\in{\{1,2,\cdots,T_{max}\}}$.
		\item The AoI vector $\boldsymbol{\varDelta}\triangleq(\varDelta_1,\varDelta_2,\cdots,\varDelta_N)$, where $\varDelta_i$ denotes the instantaneous AoI of $u_i$ at the time slot $t$.
	\end{enumerate}
	\item \textbf{Action Space:} The action $a_t\in\mathcal{A}$ contains two aspects: $i$): the power allocation policy: denote $\boldsymbol{\overset{\sim}{\alpha}}\triangleq(\alpha_1,\alpha_2,\cdots,\alpha_N)$ as the power coefficient allocated by the BS to all the $N$ users at the time slot $t$, where $\alpha_{i}\in\{\frac{1}{M},\frac{2}{M},...,\frac{M-1}{M}\}$; $ii$): the retransmission policy $\chi_i$. Noticing that when $t_i=T_{max}$, the BS will only transmit a new package for $u_i$, that is $\chi_i=0$.
	
	\item\textbf{{Reward:}} We denote the instantaneous expected weighted sum AoI as the reward $r_t$ at the time slot $t$, which can be expressed as $r_t=\sum_{i=1}^N{\omega _i\varDelta _i}$.
	
	\item {\textbf{Transition Probability:}} We denote $P\left( s_{t+1}\mid s_t,a_t \right) $ as the transition probability from state $s_t$ to state $s_{t+1}$ when taking the action $a_t$. The transition probability can be expressed as the following three parts: 
	\begin{enumerate}
		\item Power allocation coefficient:
		\begin{equation}
			\begin{cases}
\boldsymbol{\overset{\sim}{\alpha_i}}'=\boldsymbol{\overset{\sim}{\alpha_{i+1}}}\,, 1\le i\le T_{max}-2 \\
\boldsymbol{\overset{\sim}{\alpha_{T_{max}-1}}}'=\boldsymbol{\overset{\sim}{\alpha}}
			\end{cases},
		\end{equation}
	where $\boldsymbol{\overset{\sim}{\alpha_i}}'$ denotes the $i-$th power coefficient vector which is stored in the buffer at the time slot $t+1$. At each iteration, we store the BS's power allocation scheme $\boldsymbol{\overset{\sim}{\alpha}}$ in the buffer as $\boldsymbol{\overset{\sim}{\alpha_{T_{max}-1}}}$, and then update $\boldsymbol{\overset{\sim}{\alpha_i}}$ into $\boldsymbol{\overset{\sim}{\alpha_{i+1}}}$. In this way, the buffer can always maintain $T_{max}-1$ freshest power coefficient vectors for MRC decoding.

		\item AoI evolution: the transition of AoI is given as (\ref{eq:4}). To get $\varepsilon_i$ at the time slot $t+1$, we should first obtain  $\varepsilon_{ij}(t_j^{'})$, according to (\ref{eq:9}). Noticing that the MRC decoding for a $t_j^{'}$ times transmitted signal will use the power coefficient scheme $\overset{\sim}{\alpha}$ and the freshest $t_j^{'}-1$ power coefficient vectors in the buffer. Combined with (\ref{eq:8}), we can get:	
     \begin{footnotesize}
		\begin{equation}
\varepsilon _{ij}\left( t_{j}^{'} \right) =\begin{cases}
\psi _{ij}\left( \boldsymbol{\overset{\sim}{\alpha}} \right) ,&                     t_{j}^{'}=1\\
	\psi _{ij}\left( \boldsymbol{\overset{\sim}{\alpha_{T_{max}-1}}},\boldsymbol{\overset{\sim}{\alpha}} \right) ,  &             t_{j}^{'}=2\\
     \cdots\cdots &\\
	\psi _{ij}\left( \boldsymbol{\overset{\sim}{\alpha_1}},\cdots ,\boldsymbol{\overset{\sim}{\alpha_{T_{max}-1}}},\boldsymbol{\overset{\sim}{\alpha}} \right) , &  t_{j}^{'}=T_{max}\\
\end{cases}.
		\end{equation}
  \end{footnotesize}
		\item Transmission round: the transition of transmission round is given as (\ref{eq:05}).
\end{enumerate}
\end{itemize}

\section{PROPOSED DOUBLE-DUELING-DQN ALGORITHM}\label{section4}
In this section, we first describe the idea of DQN, and then utilize the Double-Dueling-DQN algorithm to solve the formulated MDP.

\subsection{Preliminaries of DQN}
The main idea of reinforcement learning (RL) is to find the optimal decision policy by maximizing the long-term discounted return. We introduce the action-value function (Q-function) to measure the value under policy $\pi$ by taking action $a_t$ in a given state $s_t$, which can be expressed as:
\begin{equation}
	\label{eq:14}
	Q^{\pi}\left( s_t,a_t \right) =\mathbb{E} _{\pi}\left[ \sum_{k=0}^{\infty}{\gamma ^k\cdot r_{t+k}|\pi} \right] \,,
\end{equation}
where $\gamma\in[0,1]$ represents the discounted factor.
Then, the value function which measures the value starting from the state $s_t$ under policy $\pi$ can be expressed as:
\begin{equation}
	V^{\pi}\left( s_t \right) =\underset{a_t}{\mathbb{E}}\left[ Q^{\pi}\left( s_t,a_t \right) \right] \,.
\end{equation}
Noticing that the Q-function evaluates the value of all actions, if we obtain $Q^*$ which is the Q-function under the optimal policy $\pi^*$, the optimal action can be selected based on the Q-value, therefore, $\pi^*$ can be obtained by:
\begin{equation}
	\pi^*=\mathop{\arg\max}\limits_{a_t\in\mathcal{A}}Q^*(s_t,a_t)\,.
\end{equation}
As a commonly used RL algorithm, Q-learning is precisely designed to obtain $Q^*$. Q-learning updates $Q^\pi$ iteratively in the following way until it converges to $Q^*$.
\begin{equation}
Q^{\pi}(s_t,a_t)=Q^{\pi}(s_t,a_t)+\lambda (y_{t} - Q^{\pi}(s_t,a_t))\,,
\end{equation}
where $y_t=r_t+\gamma \underset{a_{t+1}}{\max}Q^{\pi}(s_{t+1},a_{t+1})$ is the temporal difference target (TD target), and $\lambda$ denotes the learning rate.

However, Q-learning is difficult to converge when solving high-complexity problems. To this end, DQN algorithm combines neural networks with Q-learning to solve the high computational complexity problem. DQN is updated in a similar way to Q-learning, except that a neural network $Q^\pi(s_t,a_t;\theta)$ is used to approximate the Q-function, and the network parameter $\theta$ is updated by stochastic gradient descent method for each update.

\subsection{Double-Dueling-DQN Algorithm}
We adopt the Double-Dueling-DQN algorithm which is an improved version of DQN to deal with the problem (\ref{eq:10}). 

In an interaction, the agent observes the current transmission state $s_t$ of the HARQ-aided NOMA network and selects an action $a_t$ by $\epsilon$-\textit{Greedy Policy}. After the NOMA network takes action $a_t$, the users' AoI and the transmission state will change. Then, the NOMA network returns a reward $r_t$ and the next state $s_{t+1}$, which are in terms of $a_t$. The $\epsilon$-\textit{Greedy Policy} is given as:
\begin{equation}
	\label{eq:18}
	a_t=
	\begin{cases}
		random\,\, action,& w.p.\,\,\epsilon\\
		\mathop{\arg\max}\limits_{a}Q^\pi(s_t,a;\theta),& w.p.\,\,1-\epsilon
	\end{cases},
\end{equation}
where $\epsilon$ denotes the exploration probability. This policy makes the agent inclined to explore more actions in the early stage of the algorithm, which can avoid the algorithm from converging to a local optimum. With continuous iterations, we gradually set $\epsilon\rightarrow 1$  to obtain a stable optimal solution. 

After each round of interaction, we put $(s_t,a_t,r_t,s_{t+1})$ into the experience replay buffer $\mathcal{D}$. If the size of experiences in $\mathcal{D}$ exceeds a certain amount $\sigma$, a mini-batch experiences $\mathcal{B}=\{(s_t,a_t,r_t,s_{t+1})\}$ will be sampled from $\mathcal{{D}}$ for updating the network parameter during the training stage. This technique is called experience replay, which can improve the utilization of training data and reduce the correlation of adjacent samples.

In the training stage, to solve the overestimation problem of traditional DQN, two networks are used for training in our algorithm. The main network parameter is $\theta$, which is used to approximate the Q-function, and the target network parameter is $\theta^-$, which is used to calculate the TD target to update the main network. The TD target can be expressed as:
\begin{equation}
	\label{eq:19}
y_t^{target}=r_t+\gamma \underset{{a_t}_{+1}}{\max}Q^{\pi}(s_{t+1},a_{t+1};\theta^{-})\,.
\end{equation}
The main network updates the parameters $\theta$ by minimizing the loss function as follows at each interaction:
\begin{equation}
L(\theta)=\frac{1}{|\mathcal{B} |}\sum_{({}s_t,a_t,r_t,s_{t+1})\in \mathcal{B}}{\left[ y_t^{target}-Q\left( s_t,a_t;\theta \right) \right] ^2}\,.
\end{equation}
The target network parameter $\theta^-$ is synchronized to the main network parameter $\theta$ every $l$ updates.

Besides, we use dueling network architecture to improve the algorithm performance, which adopts two networks to approximate the Q-function. We approximate the value function $V^\pi(s_t)$ and the advantage function $A^\pi(s_t,a_t)$ with the same parameter, where $A^{\pi}\left( s_t,a_t \right) =Q^{\pi}\left( s_t,a_t \right) -V^{\pi}\left( s_t \right) $. Then, The Q-function can be obtained by these two networks, which can be expressed as:
\begin{equation}
	Q^{\pi}(s_t,a_t;\theta )=V^{\pi}(s_t;\theta )-\underset{a_t}{mean}A^\pi(s_t,a_t;\theta )\,.
\end{equation}

Details of the training process are given in Algorithm \ref{Algorithm 1}.

\begin{algorithm}
	\label{Algorithm 1}
	\caption{Double-Dueling-DQN Algorithm}
	\LinesNumbered
	\KwIn{$E,T_{slot},\epsilon,\eta,\sigma,\lambda,k,l$;}
	\KwOut{The optimal Q-network $Q^*(s,a;\theta)$;}
	Initialize the network parameter $\theta$ and $\theta-$;
	
	Initialize the state of HARQ-aided NOMA system  $s_0$;

	\For{$episode = 1$ to $E$}
	{
		Reset the system state $s_t=s_0$;
		
	\For{$t=1$ to $T_{slot}$}
		{
			Observe the system state $s_t$ and calculate the Q value of all actions $Q(s_t,\cdot;\theta)$;
			
			Select a action $a_t$ according to the $\epsilon-$\textit{greedy policy} (\ref{eq:18});
			
			Update $\epsilon =\eta \epsilon $;
			
			The BS selects action $a_t$, and gets the next state $s_{t+1}$ and the reward $r_t$;
	
			Store $(s_t,a_t,r_t,s_{t+1})$ in the replay buffer $\mathcal{D}$;
			
			\If{size of experiences $> \sigma$ and $t\,\%\, k=0$}
				{
					Sample a mini-batch $\mathcal{B}$ from $\mathcal{D}$ randomly;
					
					Calculate the TD target according to (\ref{eq:19});
					
					Training the main network with stochastic gradient descent method, and update $\theta$ with learning rate $\lambda$;
				}
			
			 \If{$t\,\%\, l=0$}
			 	{
			 		Update the target network as $\theta ^-=\theta $;
		 		}
		}
	}
	\Return{The optimal Q-network $Q^*(s,a;\theta)$}
\end{algorithm}

\section{SIMULATION RESULTS}\label{sectionVI}
In this section, we take a 4-users ($N=4$) scenario as an example to analyze the timeliness performance of the HARQ-aided NOMA network.  
We set the distance from the user to the BS as $d_1=1.5,\,d_2=2,\,d_3=2.5,\,d_4=3$, the path loss exponent as $\tau=2$, the message length as $N=160$, the number of symbols as $m=200$, and the power level as $M=20$. The training parameters are set as $E=500, \,T_{slot}=10^3,\, \epsilon=0.1,\,\eta=0.99,\,\sigma=200,\,\lambda=10^{-3},\,k=4,\,l=100$.

\begin{figure}
    \centering
   \includegraphics[angle=0,width=0.45\textwidth]{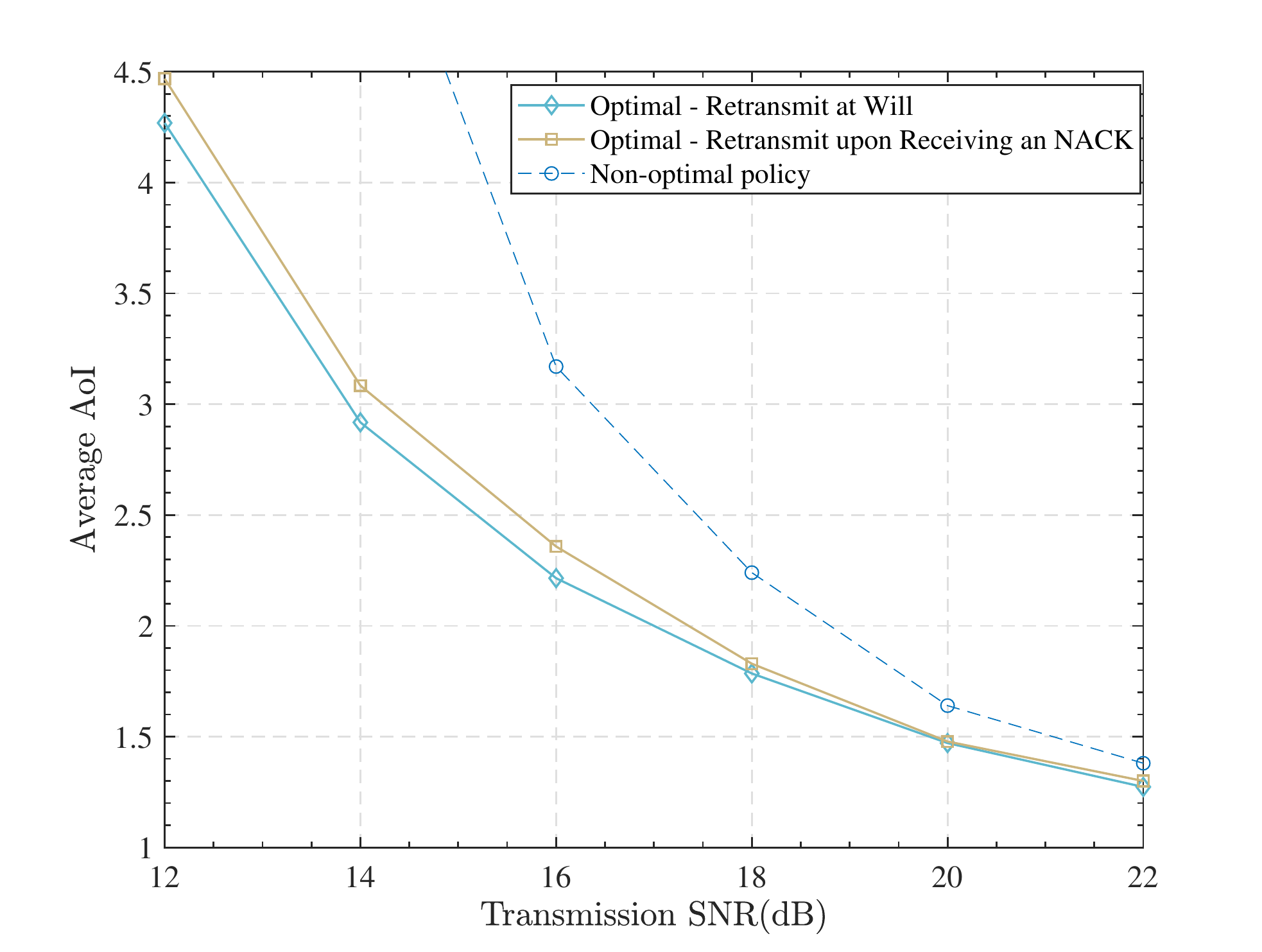}
    \caption{The average AoI under different transmission policy with maximum retransmissions $T_{max}=2$.}
    \label{fig:3}
\end{figure}

Fig. \ref{fig:3} compares the Retransmit-At-Will policy with the policy in \cite{wu2022minimizing}. Also, a non-optimal policy is given as a benchmark where the BS only transmits new packages with a fixed power allocation scheme. Retransmit-At-Will policy achieves the best AoI performance at low SNR, while two optimal policies perform similarly at high SNR. This is because retransmission is permitted when the BS received an ACK in the Retansmit-At-Will policy, which can enhance the transmission reliability at low SNR. However, since transmission is reliable enough at high SNR, the system will not choose this retransmission because it will cause stale delivery.

\begin{figure}
    \centering
    \includegraphics[angle=0,width=0.45\textwidth]{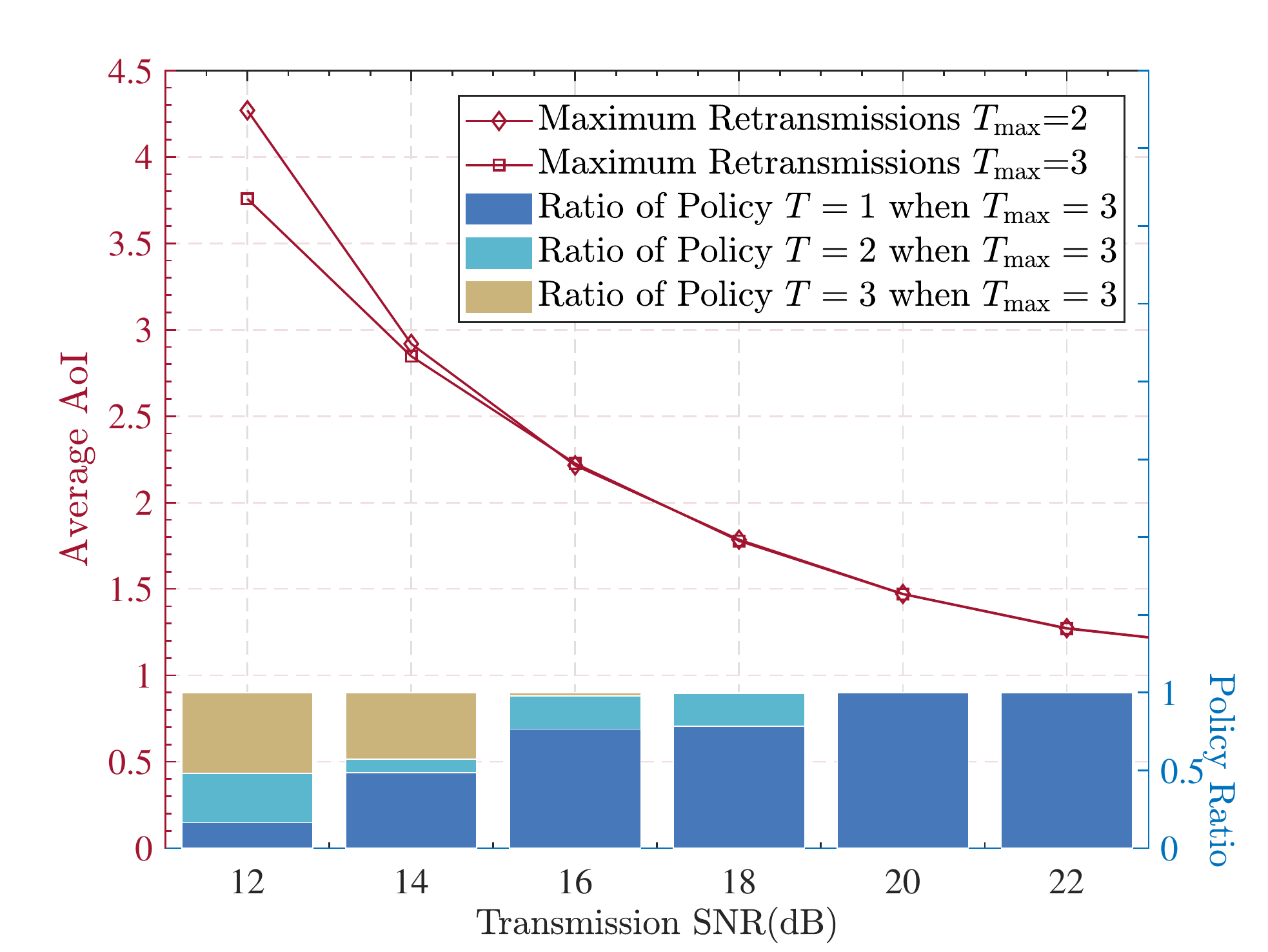}
    \caption{The average AoI under different maximum retransmissions $T_{max}$ and the policy structure when $T_{max}=3$.}
    \label{fig:4}
\end{figure}

Fig. \ref{fig:4} illustrates the AoI performance under maximum retransmissions $T_{max}=2$ and $T_{max}=3$. The network realizes better AoI performance at low SNR when $T_{max}=3$, but there is almost no difference under different $T_{max}$ at high SNR. To explain this, we demonstrate the retransmission policy when $T_{max}=3$ with the ratio of the retransmission rounds $T$. From the ratio, policy $T=3$ accounts for a larger proportion at low SNR, which means the BS prefers to take more retransmissions to ensure transmission reliability. When the SNR exceeds 16 dB, the BS prefers fewer retransmissions to ensure the timeliness of transmission, and the policy structure of $T_{max}=3$ is more like that of $T_{max}=2$, so they perform similarly at high SNR. According to this, we can find that retransmission can only improve system timeliness at low SNR. When the SNR exceeds a certain threshold, the system prefers to always transmit new packages, and retransmission is meaningless for system timeliness.

\begin{figure}
    \centering
    \includegraphics[angle=0,width=0.45\textwidth]{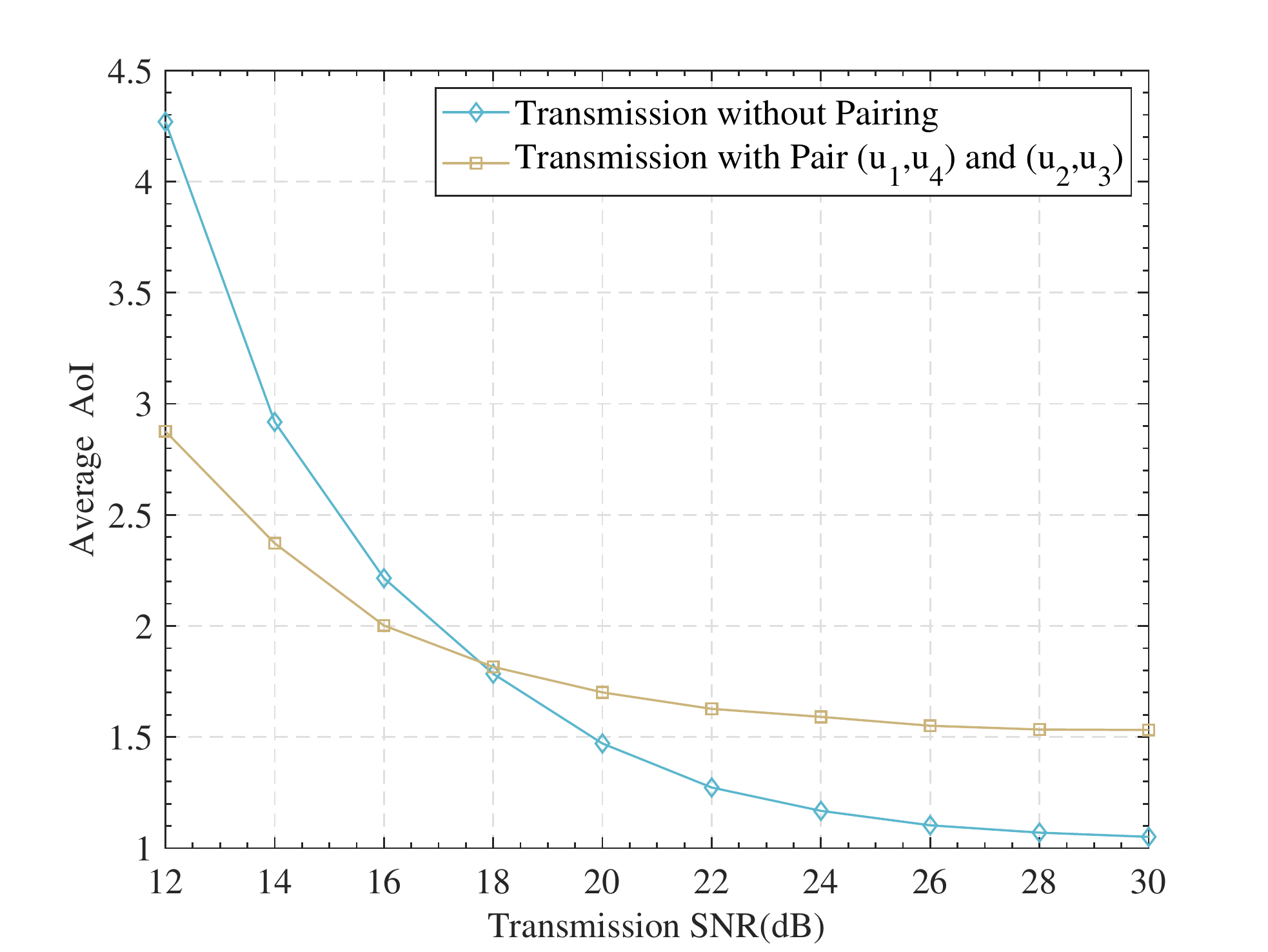}
    \caption{The average AoI performance comparison of transmission with pairing and without pairing when $T_{max}=2$.}
    \label{fig:5}
\end{figure}

To investigate the impact of user pairing on the system AoI, we set the optimal pairing scheme in \cite{8408563} as a comparing benchmark. In this case, $u_1$ is paired with $u_4$, and $u_2$ is paired with $u_3$. The BS only transmits packages for one user pair at a time slot, and the other user pair can only wait at this time. Fig. \ref{fig:5} compares the performance of the Retransmit-At-Will policy with and without user pairing. We can find that transmission with pairing has a better AoI performance at low SNR, while transmission without pairing performs better at high SNR. That is because the user pairing technique reduces the users served simultaneously, which can reduce the complexity of SIC and ensure transmission reliability at low SNR. While reliability can be assured at high SNR, thus transmission without pairing performs better because it can serve more users simultaneously and save waiting time. As a result, user pairing can be used to enhance the timeliness at low SNR.

\section{CONCLUSION}
In this paper, we optimize the AoI of the HARQ-aided NOMA network with multiple users and unrestricted retransmissions, which can intelligently decide how to allocate power resources and when to retransmit. First, a more flexible retransmission scheme with better AoI performance, referred to as Retransmit-At-Will is proposed. Then, we reformulate the age-optimal problem as an MDP, and the Double-Dueling-DQN is adopted to obtain the optimal policy. At last, a threshold structure of the retransmission policy is demonstrated, and user pairing is proved to be a technique enhancing the system freshness at low SNR.

\bibliographystyle{IEEEtran}
\bibliography{reference}

\end{document}